# PRIORITY BASED BANDWIDTH ALLOCATION IN WIRELESS SENSOR NETWORKS

Mary Cherian[1] and T.R. Gopalakrishnan Nair[2]

[1]Department of Computer Science & Engineering, Dr.Ambedkar Institute of Technology, Bangalore, India
[2]Dayananda Sagar Institutions, Bangalore, India

## ABSTRACT

*Most of the sensor network applications need real time communication and the need for deadline aware real time communication is becoming eminent in these applications. These applications have different dead line requirements also. The real time applications of wireless sensor networks are bandwidth sensitive and need higher share of bandwidth for higher priority data to meet the dead line requirements. In this paper we focus on the MAC layer modifications to meet the real time requirements of different priority data. Bandwidth partitioning among different priority transmissions is implemented through MAC layer modifications. The MAC layer implements a queuing model that supports lower transfer rate for lower priority packets and higher transfer rate for real time packets with higher priority, minimizing the end to end delay. The performance of the algorithm is evaluated with varying node distribution.*

## 1. INTRODUCTION

A wireless sensor network (WSN) comprises of sensor nodes, which monitor the terrain where they are deployed and gather the physical environmental parameters which they communicate to the base station [1]. The sensor nodes are densely deployed in regions where they monitor many physical phenomena such as vibration, movement of objects, temperature, humidity, pressure, radiations, noise levels, and light conditions. The sensor nodes self-organize to form an Ad-hoc network after the deployment. The sensor nodes are resource constraint as these are equipped with batteries with limited power, tiny microprocessors/microcontrollers, low power transceivers, and sensors for gathering information about the deployed environment. When a single sensor node is limited in its capabilities, the composition of large number of nodes offers technological capabilities. In wireless sensor networks, individual sensor nodes are inherently unreliable and have very limited capabilities to ensure real-time guarantees. The target is to provide more reliable services with reduced end-to-end delays, and lower energy consumption in the underlying sensor network. A hard real-time (HRT) system guarantees that critical tasks complete on time [2], [3]. Here meeting a deadline is mandatory otherwise system failure occurs which may cause catastrophic damage [4]. These systems are implemented as safety critical systems such as in aerospace and defence. Performance of such systems cannot be compensated over any other feature of the system. A certain amount of latency is allowed for soft-real time solutions. In such cases the transferred data is not critical and the system used may use the soft real time solutions for directing the information to the sink. These systems are known as non-safety critical systems where the system deals with non critical data and use soft real time solutions for data transfer.

The applications of wireless sensor networks spread in different domains viz. military, emergency situation management, physical world, medical and health, industry, home network and automotive. Most of the sensor network applications need real time communication and the need for deadline aware real time communication is becoming eminent in these applications. These applications have different dead line requirements also. Some of the challenges for real time communication are random deployment, dynamic network topology, traffic characteristics, resource constraints, and transient congestion.

In WSN the sensed data from the nodes may travel multiple paths towards the destination. The available bandwidth in the channel will be portioned, and the response times will become very lengthy [5]. The real time applications of wireless sensor networks are bandwidth sensitive and needs higher share of bandwidth for higher priority data to meet the dead line requirements. Through the assistance from MAC layer, the availability of multiple non-interfering and prioritized paths can be made available to the routing algorithm. In this paper we focus on the MAC layer modifications to achieve the real time requirements of the different priority data. In the layered view of a network model the MAC layer should guarantee the channel access delay. The network layer routing protocol should bind the end-to-end transmission time [6].

The remainder of the paper is organized as follows. Section 2 aims to provide a survey on the state of the art of related MAC and routing protocols. Section 3 provides an overview of the system. The outcomes of the implemented technique are discussed in section 4. Finally, section 5 contains some concluding remarks and the potential research directions.

## 2. RELATED WORK

In WSN the MAC layer has an important role in deciding the channel access delay which contributes to the end-to end delay. Contention based protocols [6] can reduce the collisions, but a deterministic channel access cannot be decided. On the contrary in TDMA based MAC a bounded and predictable medium delay can be determined, but a central co-ordination is required [6]. The original IEEE 801.11 [7] uses DCF (Distributed Coordination Function) and PCF (Point Coordination Function). CSMA/CA (Carrier Sense Multiple Access with Collision Avoidance is used for the channel access. If channel is free, it waits for DIFS (Distributed inter frame space) and transmits if still free. If channel is busy, it randomly backs off a number of Slots (minimum 15, maximum 1023). If packet transmission further fails, back-off is increased to a random window, up to a preconfigured upper limit. But the access method neither differentiates the different priority data nor reserves the bandwidth. PCF is centrally controlled and uses polling. The PC (Point Coordinator) maintains the list of nodes eligible for polling. As PCF is not compatible with all type of real time data, it is not practically used.

The protocols described in [6], S-MAC, T-MAC, B-MAC apply CSMA/CA for medium access, but they only provide best effort service. 802.11e [8], [9] provides QoS support for wireless networks at the MAC layer of the protocol stack and enhances the medium access functions. To achieve hard time data delivery of the packets in addition to a HRT MAC layer, a hard real time routing protocol is also essential. Taimur Qureshi et al. in [5] present a network Layer based Hard Real Time Protocol for wireless sensor networks. But the protocol assumes a MAC (Medium Access Control) layer which is collision free and with, known medium access delay. It also does not consider the priority of the packets. Other existing real time routing protocols for WSN viz. RAP [10], SPEED [11], R2TP [12] do not have a priority based data delivery mechanism. They neither consider a queuing model for packet scheduling at the nodes nor allocate the bandwidth based on the priority of the packets.

We have presented a real time routing protocol, RRDTE in [13] which detects time critical events in WSN. But the bandwidth allocation is not addressed in this protocol. In [14], we propose a scheduling algorithm for sensor networks based on the packet service ratio which helps in congestion control and flow control. But this also does not address the bandwidth partitioning based on the priority of the packets. M.Caccamo et al., in [15] propose an implicit prioritized access protocol, for deterministic real time communication which demands prior topological information and synchronization. However, this is not suitable in non-deterministic environments. In I-EDF nodes are organized in hexagonal cells. Each cell will have a unique frequency in intra-cellular communication. The six directions of the hexagon are numbered and the communication slots alternate with a given direction in inter-cellular communication. The rigid cell-based organization of the topology has a limitation in real environment where there is random deployment of sensor nodes.

Watteyne et al. in [16] present a novel hard-real time MAC protocol with realistic assumptions. The focus is mainly on the MAC layer, as hard real-time needs to be dealt at each layer in communication architectures. The protocol consists of an initialization phase, followed by a run-time phase, subdivided in two modes: unprotected and protected mode. Unprotected mode is used when collision probability is low and multi-hop propagation speed is near optimal. When collision occurs, the network switches to a slower but collision-free protected mode. But the protocol works only on linear topologies. This protocol also requires dedicated frequency channels. DMAC [6], DR-MAC [6] protocols minimize the communication latency, but they are suitable only for specific tree topologies.

CE. Perkins et al. in [20] presents the AODV, a routing protocol for Ad-Hoc networks, but priority based data delivery is not addressed in this protocol also. Iftikhar Ahmad et al. in [21] present a routing protocol by improvising the basic routing mechanism in AODV to assign more bandwidth to real time traffic (RT), compared to best effort traffic (BE). The protocol reserves a larger time slot for transmission of RT traffic than that of BE traffic. The protocol assumes only one queue for the buffering of the packets. Here the BE traffic may be starving and may be dropped also, if the queue fills up.

## 3. OVERVIEW OF THE SYSTEM

Sensor nodes share the available resources such as transmission bandwidth, buffer storage and the processing capability. The delay and loss performance can be quantified as explained in this subsection. When a packet or connection request arrives at a node, the node may be in a blocking state because of the unavailability of the resources. The data packets arrive at a node in a random manner; the time that they spend in the node is also random. Due to lack of resources, the packets may get blocked or lost. Throughput of the system is the long-term departure rate from the system. The average arrival rate is given by the inverse of the average inter arrival time. Little's formula[17] can be applied to a networking scenario also in which the system can be a network node, transmission line, multiplexer, a switch or a network itself [18].

### 3.1. Priority queuing at the node

A queuing model is considered at the nodes where requests can belong to one of N priority classes. When a packet arrives at the system, it joins the queue of its priority class. Each time one packet is fetched, from the queue, the next packet to be served is selected from the head of the line of the highest priority non empty queue. It is assumed that the arrival at each priority class is Poisson with rate $i_n$ and that the average service time of a class N customers is E $[s_n]$. So the load offered by class N is, $n = i_n$ E $[s_n]$ where is the utilization factor.

The highest priority class (class 0) has average waiting time

$$E\,[W0] = (i_n\,E\,[s^2]) \diagup (1 - 0) \tag{1}$$

So E [W0] saturates as approaches 1. Thus the saturation point of class 0 is determined only by its own load. On the other hand, the waiting time for class 1 is given
By

$$E\,[W1] = (i_n\,E\,[s^2]) \diagup (1 - 0)\,(1 - 0 - 1) \tag{2}$$

The class 1 queue will saturate when 0 + 1 approaches 1. Thus the class 1 queue saturation point is affected by the class 0 load. Similarly, the class N queue saturation point depends on the sum of the loads of the classes of priority up to N.

At the MAC layer Enhanced Distributed Channel Access (EDCA) method is used. Advantage here is that this allows 'tuning' of the parameters for any specific network/application. Four priority categories are Class 0, Class 1, Class 2, and Class 3 with Class 0 having the highest priority. Each Access Category is equipped with a single transmit queue as depicted in Figure 1.

Table 1 Packet Priority Classes

| Class | Priority | Traffic Type |
|-------|----------|--------------|
| 3 | Priority 3 | Normal- low priority periodic data |
| 2 | Priority 2 | Control data |
| 1 | Priority 1 | Real-Time  data |
| 0 | Priority 0 | Real-Time critical data(Lowest Delay) |

The MAC layer service data unit (MSDU) with UP (user priorities) which is transferred from the higher layer (network layer) is mapped, and subsequently placed in the corresponding priority queue. This ensures bandwidth partitioning among different priority transmissions.

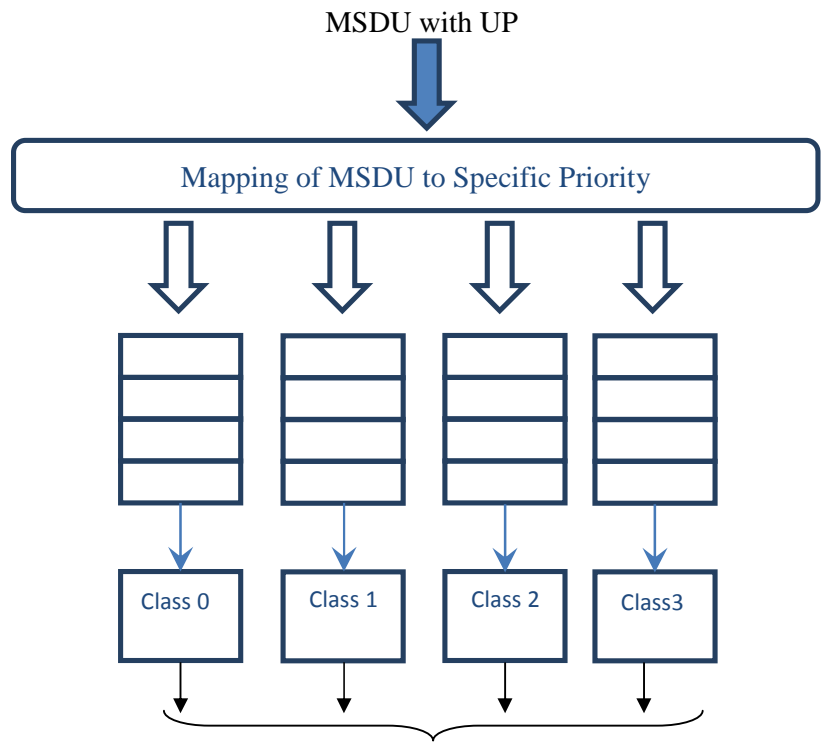

Figure 1. Queues with Different Priorities

The routing protocol applied is the hard real time protocol for wireless sensor networks (RPS) which is a multipath routing protocol. Protocol discovers disjoint paths. However, the RPS algorithm is based on the computation of dynamic routes and can handle the dynamic changes of the networks. Delay as the metric, a source node selects its route dynamically and checks the quality of the alternative routes.

## 4. SIMULATION RESULTS AND DISCUSSIONS

In this work, we consider the bandwidth allocation based on priority of the data for the applications where sensor nodes are deployed in Ad-hoc manner to detect critical events. We have applied the discrete event simulator (NS-2) [19] to model the behaviour of the application. The nature of the events generated in the system is stochastic, discrete, and dynamic.

A single base station gathers data from the sensor nodes. The simulations are conducted with 32, 64,128, and 256 nodes. The wireless sensor nodes are deployed with random topology in terrains ranging in sizes from 500meter x 500meter to 1500meter x 1500meter. The enhanced 802.11e EDCA is employed at the MAC layer. Traffic with different priorities is generated by the application. The bandwidth allocated to each priority data is estimated based on the simulation results. The results of simulation with NS-2 simulator are charted below.

Figurers 2 to 5 indicate the bandwidth allocations for different priority data for node distributions for 32, 64, 128, and 256 nodes respectively. Figure 2 depicts the simulation results with the number of sensor nodes as 256. The Class 0 (Priority 0) traffic with the highest priority (Table 1) gets the highest share of bandwidth and the Class 3 (Priority 3) traffic gets the lowest share of bandwidth. This in turn ensures that the highest priority packets are delivered with minimum delay meeting the critical data delivery requirements.

Figure 2. Bandwidth allocations for different priority data for 256 nodes

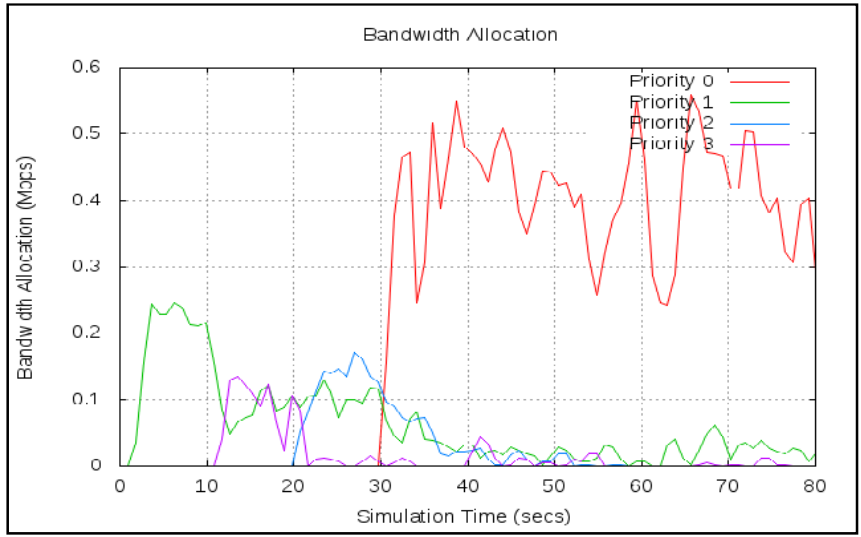

Figure 3. Bandwidth allocations for different priority data for 128 nodes

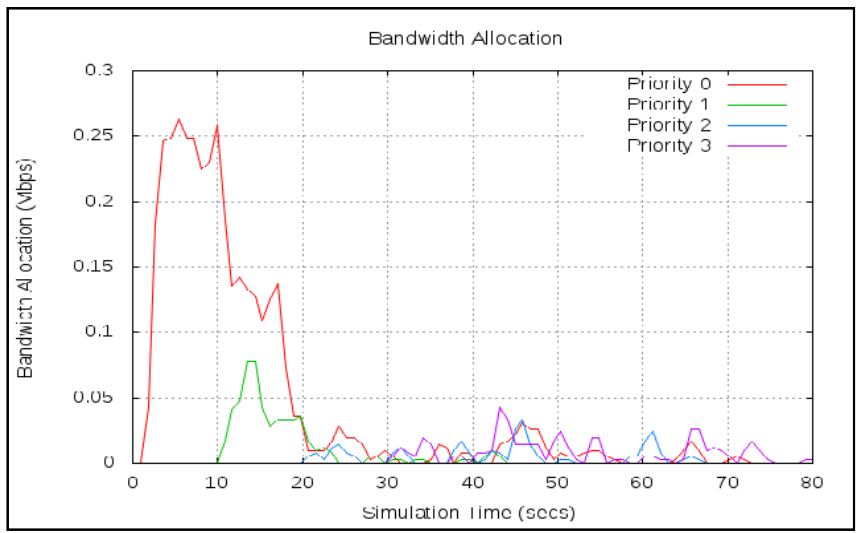

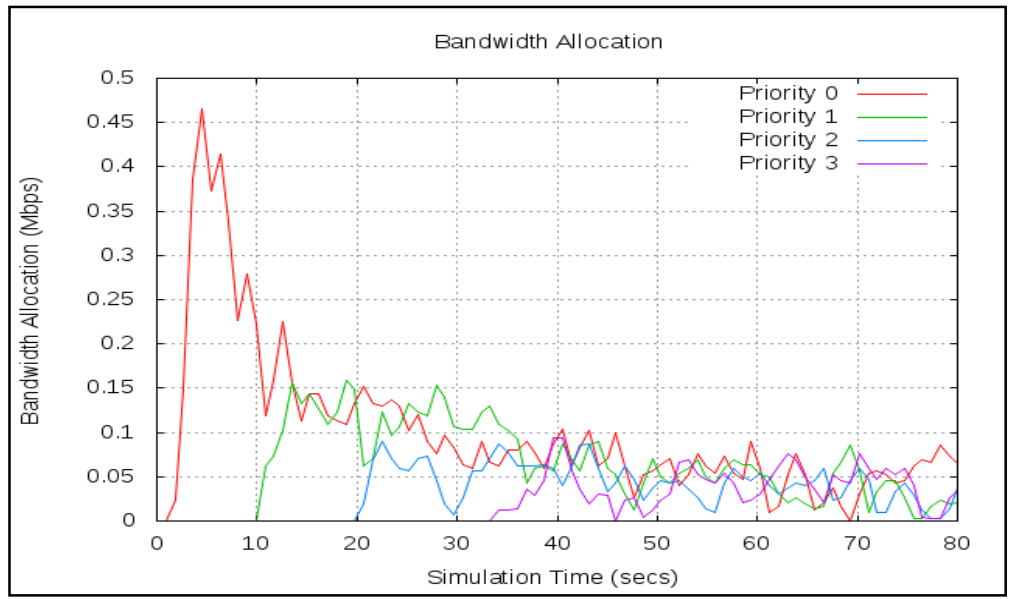

Figure 4. Bandwidth allocations for different priority data for 64 nodes

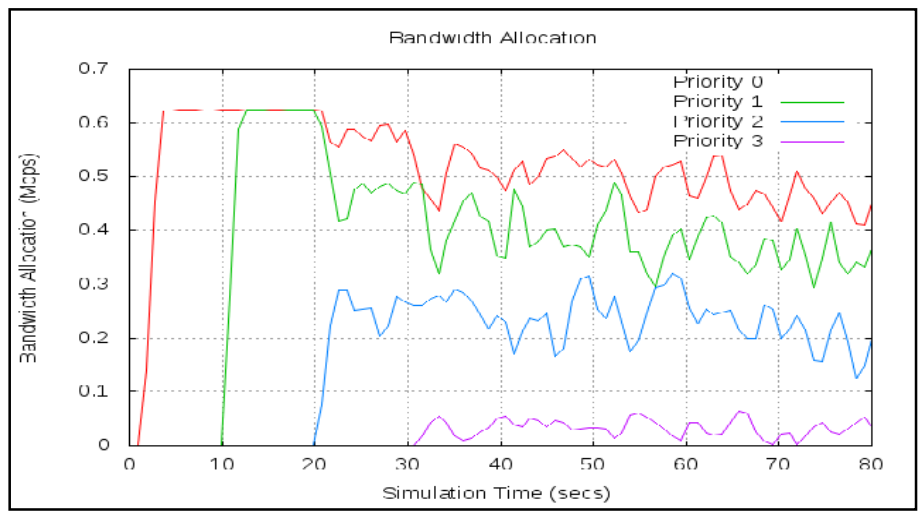

Figure 5. Bandwidth allocations for different priority data for 32 nodes

Figures 3, 4, 5 indicate the results of bandwidth allocation with 128, 64 and 32 nodes respectively. The simulation results with 128, 64 and 32 nodes are consistent with the result of 256 nodes. The percentage of bandwidth allocated for higher priority data is higher. The critical data which is event driven data [22], can be delivered with minimum delay meeting the requirements of the application.

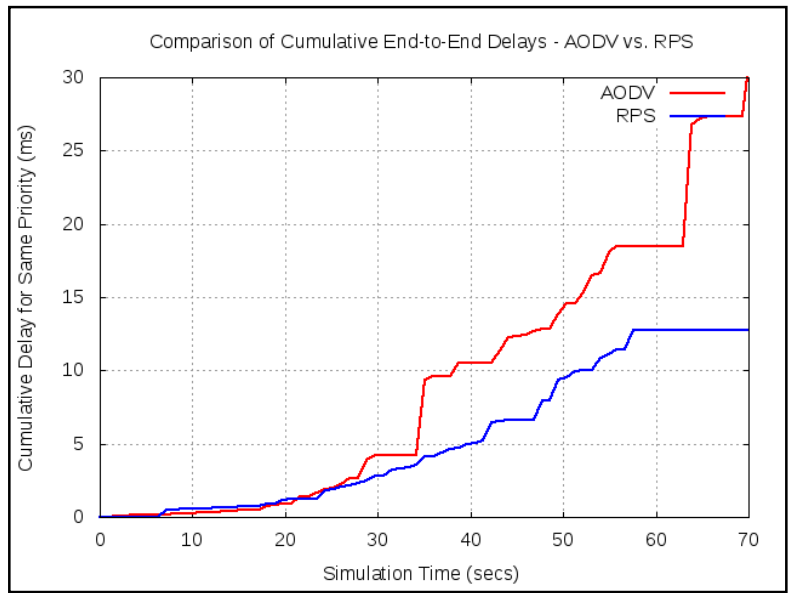

Figure 6. End-to-End Delay

In figure 6 we are comparing the end-to end delay of packet delivery in our method with the end-to end delay incurred in transmission of data packets with the AODV [20] protocol for priority 0, the critical data. AODV is an on-demand routing protocol which is widely used for wireless networks. This protocol has less routing and computational overheads, simplicity, and good scalability. But the protocol does not consider QoS or reliability as demanded by the traffic [21]. AODV does not differentiate the packets with different priority. AODV applies random packet scheduling [23]. The end-to end delay incurred in our protocol is less than that of AODV protocol as indicated by the graph in figure 6. The real time critical data is assigned the highest share of bandwith and hence is delivered with minimum dealy.

## 5. CONCLUSION

This paper details the priority based data delivery to the sink node from the sensor nodes. In this work, we consider the bandwidth allocation based on the priority of the data for the applications where sensor nodes are deployed in Ad-hoc manner to detect critical events. The data generated in WSN has different priority levels. WSNs are application specific and several applications of WSN are designed for vital event monitoring and to ensure timeliness and reliability for the measured environmental values [13]. The real time data has to be prioritized based on the urgency and resources such as bandwidth needs to be allocated accordingly to ensure the timely delivery of the data to the sink node. The bandwidth allocation for each flow depends on the QoS requirement of the specific flow [24]. The MAC layer implements priority based medium access, and queuing mechanism, which ensures the critical data delivery in a prioritized manner with bandwidth partitioning among different priority transmissions. We have applied the discrete event simulator (NS-2) to model the behavior of the system. Simulation has been conducted for different node distributions. The results are compared with similar known technique. Our technique reduces end-to end delay in data delivery as depicted in figure 6. Future research direction of this work could be an enhancement to support load balancing. We have considered a static base station and nodes in this work. Studies may be further extended for scenarios considering the mobility of the sink, the sensor nodes or both.

**AUTHORS**

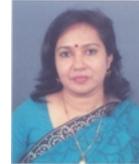

Mary Cherian has 30 years of experience in professional field spread over education, research and industry. She holds B.E degree in electronics and communication from Kerala University, India and M.Tech. in computer science and engineering from Visvesvaraya technological university Belgaum, India. She started her career in 1984 as a research engineer in O/E/N India and has worked in the field of engineering and software in industries like Kerala state electronics Development Corporation, Keltron controls, Electronics Research and Development Centre, ABB and Chemtrols Software private limited    in capacities of System Engineer, System Manager and Director. Later, she concentrated on education and contributed in academic field for the last 15 years in India and abroad educating pupil in the field of science and technology especially in Computer Science and Engineering. Currently she is working as Associate Professor in the department of Computer Science and Engineering in Dr. Ambedkar Institute of Technology, Bangalore, India. She has publications in national and international conferences and journals. Her areas of interests include Computer networks, Sensor Networks, Real time routing protocols, and Cognitive routing. Prof Cherian has membership of professional bodies such as CSI, ISTE, IE and IETE.

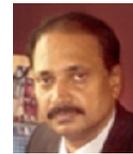

Dr. T. R. Gopalakrishnan Nair  has 35 years of experience in professional field spread over education, research and industry. He holds degrees M.Tech. from I.I.Sc., Bangalore, India and  Ph.D. in computer Science, from Kerala University, India. He started his career in Electronics Research and Development Centre, Trivandrum, India where he was instrumental in developing various pioneering research products in the field of computers and software. Later, as the head of Advanced Simulation Activities in Indian Space Research Organization, his areas of research were Critical Real-Time Systems, Inertial Navigation and Guidance systems, Launch Vehicle Technology, High speed computing and Launch Vehicle simulations. Later, he concentrated on education and contributed extensively in academic field for the last 14 years in India and abroad educating pupil in the field of science and technology especially in Computer Science and Engineering. He authored and published about 110 papers in these multidisciplinary fields and he promotes cross domain fusion of knowledge.  He has authored several book chapters at international levels and delivered keynote and invited lectures. He is the Chief Editor of Journals "Inter JRI Science and Technology and Inter JRI Computer Science and Networking" published by Interline Publishers. In 1992, he received the National Technology award PARAM from the advisor to Prime Minister, for developing the parallel computing flight simulation systems. He received the Team Tech Foundation Award for 'Excellence in Education and Research' in 2009.  Dr. Nair is a senior member of IEEE for last two decades and a member of various other societies like ACM. His areas of interests include Computer networks, Cognitive routing, Software Engineering, Bio-Informatics, AI & Robotics and Signal and Image Processing.